\documentstyle[11pt,aaspp4,epsf]{article}  
\newcommand{\designercut}{``Designer Cut''}
\received{31 Oct 1996}
\revised{13 Oct 1997}
\accepted{14 Oct 1997}
\newcommand{\be}{\begin{equation}}
\newcommand{\ee}{\end{equation}}
\newcommand{\bc}{\begin{center}}
\newcommand{\ec}{\end{center}}
\newcommand{\etal}{{\it et al.}}
\newcommand{\bea}{\begin{eqnarray}}
\newcommand{\eea}{\end{eqnarray}}
\newcommand{\COBE}{{\sl COBE}}

\slugcomment{accepted by the ApJ on 14 October 1997}

\lefthead{Wright}
\righthead{DIRBE Angular Power Spectra}

\begin{document}

\title{Angular Power Spectra of the \COBE\/ DIRBE Maps}
\author{Edward L. Wright}
\affil{UCLA Dept. of Physics \& Astronomy}
\authoraddr{Division of Astronomy\\P.O. Box 951562\\
Los Angeles CA 90095-1562}

\begin{abstract}
The angular power spectra of the infrared maps obtained by the DIRBE
(Diffuse InfraRed Background Experiment)
instrument on the \COBE\footnotemark[1]\ satellite 
have been obtained by two methods: the Hauser-Peebles
method previously applied to the DMR maps, and by Fourier transforming
portions of the all-sky maps projected onto a plane.  The two methods
give consistent results, and the power spectrum of the high-latitude
dust emission is $C_\ell \propto \ell^{-3}$ in the range
$2 < \ell < 300$.
\end{abstract}

\footnotetext[1]{The National Aeronautics and Space Administration/Goddard
Space Flight Center (NASA/GSFC) is responsible for the design, development,
and operation of the Cosmic Background Explorer (\COBE).
Scientific guidance is provided by the \COBE\ Science Working Group.
GSFC is also responsible for the development of the analysis software
and for the production of the mission data sets.}

\keywords{ISM: structure; dust; cosmic microwave background}

\section{Introduction}

The angular power spectrum of the sky brightness at millimeter wavelengths
can be used to determine the parameters of cosmological models, such
as $H_\circ,\;\Omega_\circ,\;\Omega_B$ and $\Lambda$ 
(Jungman \etal\markcite{JKKS96} 1996),
but our ability to measure this power spectrum will depend on the
angular power spectrum of foreground emission from the Milky Way.
Kogut \etal\markcite{KBBGR96} (1996) have found that free-free emission, 
which will be the dominant foreground contamination, is better correlated
with far infrared dust emission than with synchrotron emission.
The angular power spectrum of the dust emission also
determines the beam size dependence of the confusion limit to the 
sensitivity of point source surveys.
Gautier \etal\ \markcite{GBPP92} (1992) found a power spectrum
$P(k) \propto k^{-3}$ for the IRAS 100 $\mu$m maps,
in a range of $k$ corresponding to $200 < \ell < 3000$,
and Kogut \etal \markcite{KBBGR96} (1996) found $C_\ell \propto \ell^{-3}$ 
for the DIRBE 
(Boggess \etal\markcite{BMWBC92} 1992)
240 $\mu$m map in the $2 < \ell \leq 30$.
% referee comment `p3.1'
But the DIRBE instrument has an angular resolution of $0.7^\circ$, which allows
one to measure the power spectrum to angular frequencies much higher than
$\ell = 30$, and DIRBE has 10 bands (240, 140, 100, 60, 25, 12, 5, 3.5, 2.2 and
1.6 $\mu$m) which allow one to measure the frequency dependence of the angular
power spectrum.
In this note I use a simple method to allow for the effect of detector
noise, which dominates the $100 < \ell$ portion of the DIRBE 240 $\mu$m
power spectrum, and show that the 60--240 $\mu$m DIRBE maps
follow a $C_\ell \propto \ell^{-3}$ spectrum over the range 
$2 < \ell < 300$.
Since this spectrum is more than a full power of $\ell$ steeper than
the predicted spectrum of the cosmic microwave background in
inflationary dark-matter dominated models, observations of the ``Doppler''
peak at $\ell \approx 200$ should be much less affected by Galaxy than
the {\sl COBE\/} DMR maps.

\section{Method}

\begin{figure}[tbp]
\plotone{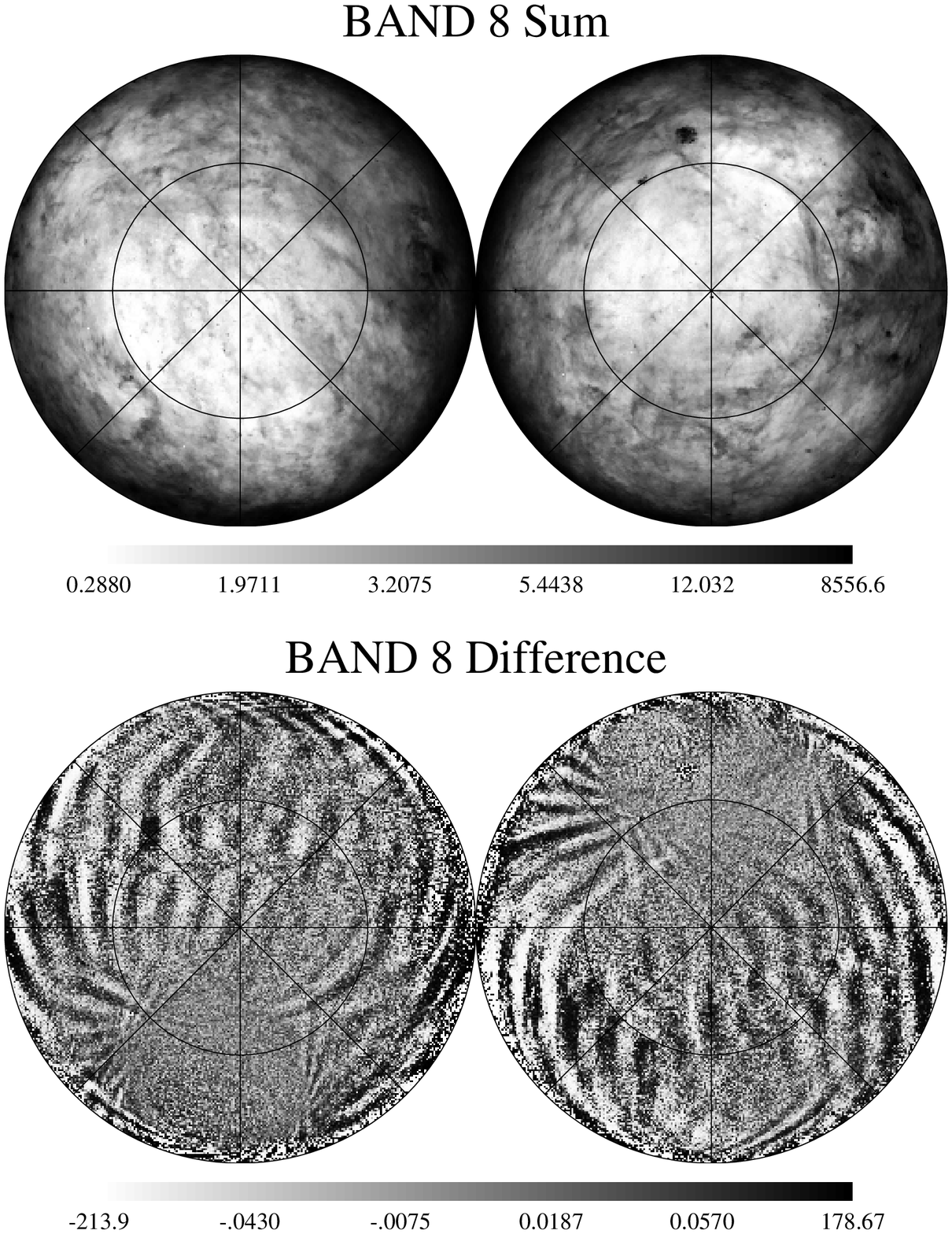}
\caption{Top: 100 $\mu$m map from the sum of the
DIRBE even and odd week sum maps.  The North Galactic Pole is the
center of the left circle, while the South Galactic Pole is the center
of the right circle.
Bottom: Difference of the even and odd sum maps.\label{fig:band8sd.eps}}
\end{figure}

\begin{figure}[tbp]
\plotone{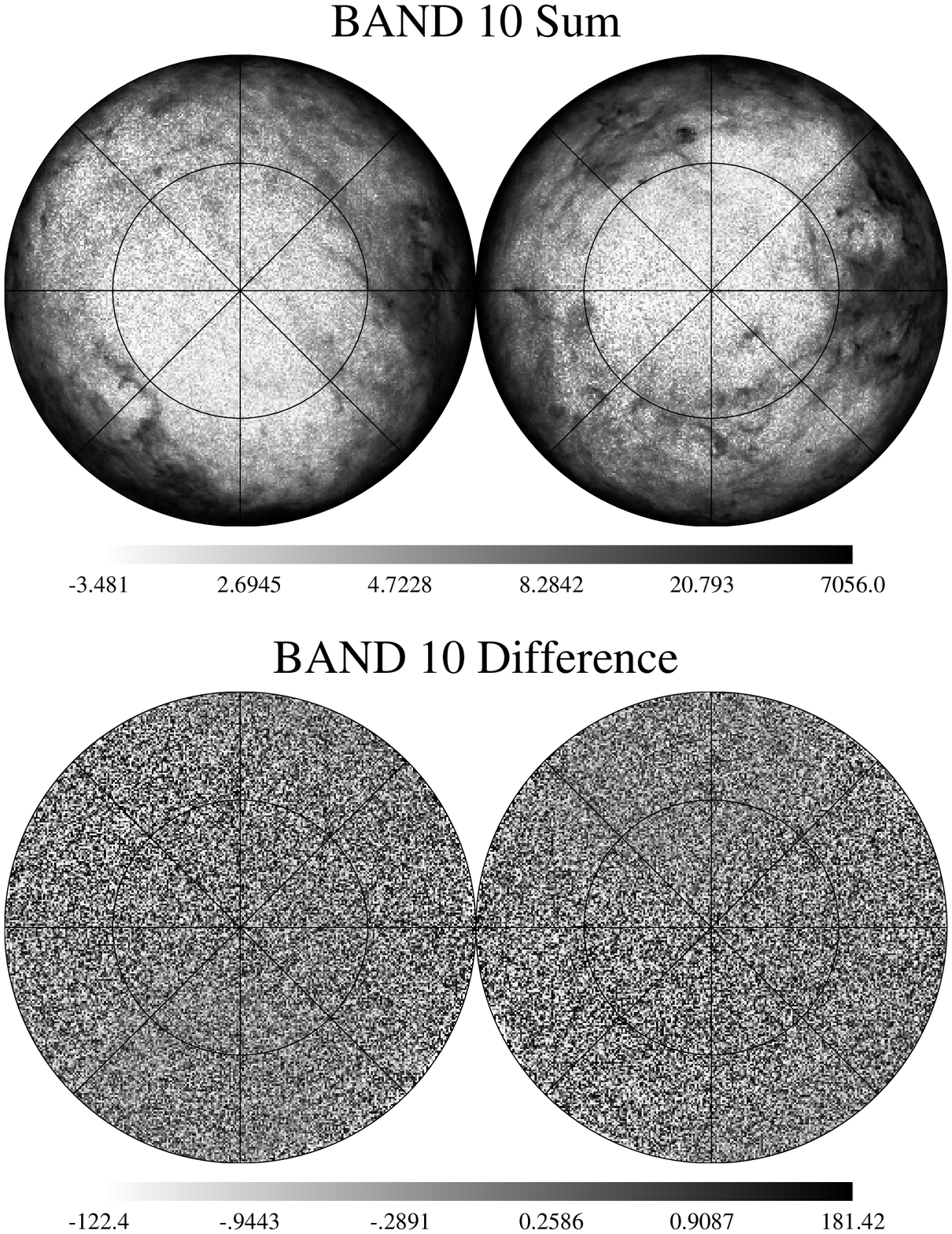}
\caption{Top: 240 $\mu$m map from the sum of the
DIRBE even and odd week sum maps.
Bottom: Difference of the even and odd sum maps.\label{fig:band10sd.eps}}
\end{figure}

The power spectrum of a process is a quadratic function of the data,
so noise will bias the power spectrum even if it is not correlated
with the signals.  The technique used by 
Wright \etal \markcite{WSBL94} (1994) to correct for this effect in the
{\sl COBE} DMR maps was to use the two sides of the DMR instrument,
the $A$ and $B$ sides, to make sum and difference maps,
$S = (A+B)/2$ and $D = (A-B)/2$.  
% referee comment `p3.2'
The noise power spectrum canceled
out when the difference in the power spectra of the $S$ and $D$ maps was taken
to yield the the power spectrum of the sky.
The DIRBE experiment does not have a comparable duplication of detectors,
but I have computed maps with independent detector noises by forming
sums of the even weeks and the odd weeks of the data.  The DIRBE
data have been averaged into separate files for each of the 41 weeks of
data taking to allow for the removal of the time varying emission from
interplanetary dust.
% referee comment `p4.1'
This time variation is due to the motion of the Earth around the Sun, which
means that a given spot on the sky is viewed through different parts of the
interplanetary dust cloud, and the time variation is used to determine the
properties of the interplanetary dust.
Once the zodiacal light is removed
(see Appendix \ref{app.zodi}), the weeks are
summed into all-sky, all-data maps.  By summing the even and odd weeks,
I make $E$ and $O$ maps, and construct the sum and difference maps as
$S = (E+O)/2$ and $D = (E-O)/2$.  
Figure \ref{fig:band8sd.eps} shows the sum and difference
maps for 100 $\mu$m DIRBE maps, while Figure \ref{fig:band10sd.eps} 
shows the 240 $\mu$m maps.  The pass 2B DIRBE weekly maps, available since
1994 on the NSSDC, have been used in this paper.
% referee comment `p4.2'
Figures \ref{fig:band8sd.eps} and
\ref{fig:band10sd.eps} have been histogram equalized, so the values on the
scale bar correspond to the $0^{th}$, $20^{th}$, $40^{th}$, $60^{th}$, 
$80^{th}$ and $100^{th}$ percentiles of the map.
The difference map at 240 $\mu$m is clearly noise dominated,
with the middle 60\% of the histogram spanning 1.85 MJy/sr.
A Gaussian distribution with a standard deviation of 1.1 MJy/sr would have
the same middle 60\% span, but the actual distribution of values is not 
Gaussian
due to variable coverage and to errors caused by large sky gradients near the
galactic plane.
This detector noise is a very significant contribution to the power
spectrum of the sum map.  
The difference map at 100 $\mu$m is dominated
by a
% referee comment `p4.3'
systematic
error in the zodiacal light removal caused by an inadequate model for the
interplanetary dust cloud,
but the overall middle 60\% span of 100 kJy/sr is
very small compared to the sum map.  
The detector noise in the 100 $\mu$m channel on DIRBE is 100 times smaller
than the detector noise in the 240 $\mu$m channel 
(Boggess \etal\markcite{BMWBC92} 1992) but the middle 60\% span of the
difference map is only 19 times smaller, showing that the 100 $\mu$m difference
map is not detector noise dominated.
Since the zodiacal light model error
has a period of two weeks due to the alternating signs given to successive
weeks in the difference map, the dominant spherical harmonic in this
difference map is $\ell = 26$.  The sum map will not have this $\ell = 26$
power because it does not apply alternating signs to successive
weeks, so this method of correcting for noise bias will not work for
maps where the zodiacal light errors are comparable to or larger than
the true signal.  This limits my analysis to the $\lambda \geq 60\;\mu$m 
DIRBE maps.  
% referee comment `p4.4'
The 25 and 12 $\mu$m DIRBE maps are so dominated by the
interplanetary dust emission that the residuals due to inadequate models,
while only a few percent of the peak emission, are still larger than the
high galactic latitude emission from interstellar dust.

\section{Variation with latitude cutoff}

\begin{figure}[tb]
\plotone{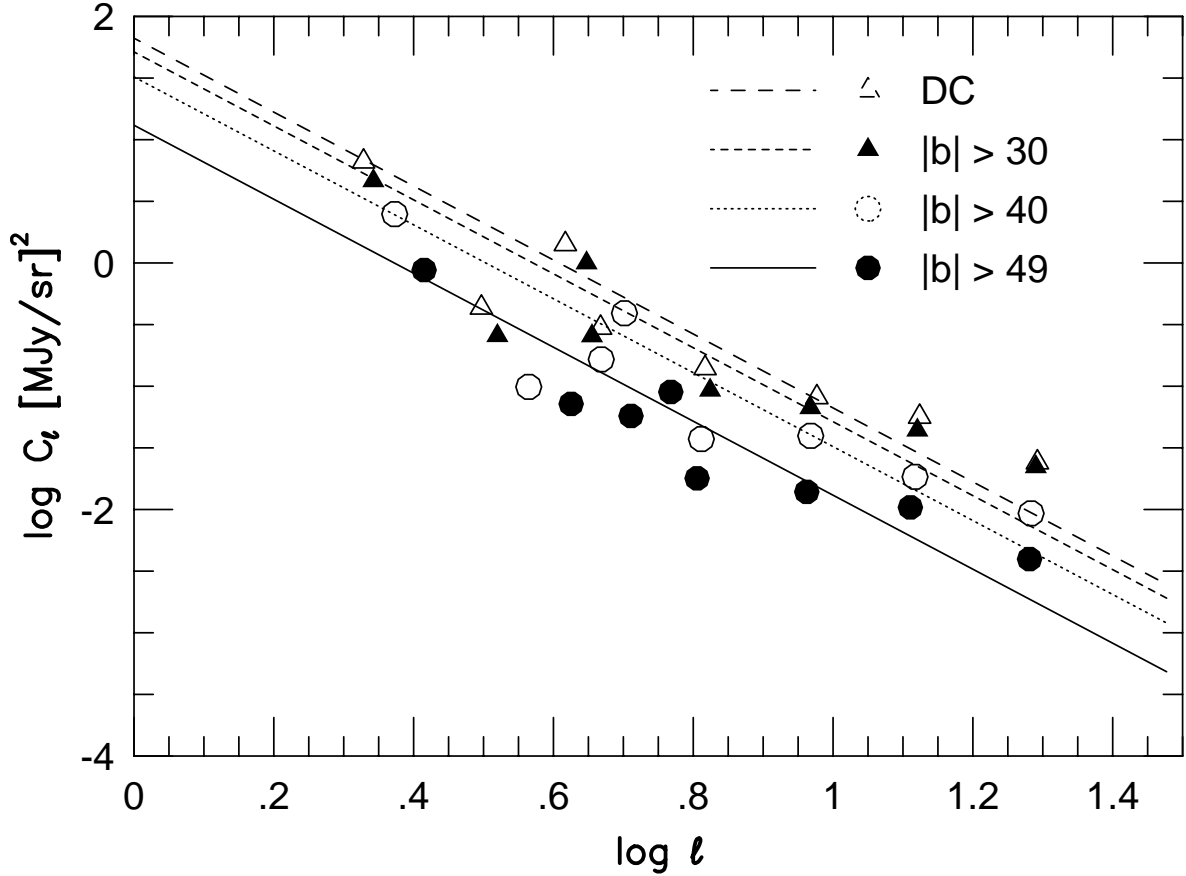}
\caption{Power spectra of the DIRBE 240 $\mu$m
map after 4 different galactic plane cuts: the \designercut,
and $|b| > 30^\circ, \; 40^\circ$ and $49^\circ$.
% referee comment `Fig 3'
The lines are best fits with an assumed $\ell^{-3}$ slope.
\label{fig:c_l_vs_b.eps}}
\end{figure}

The power spectrum of the DIRBE maps is a strong function of the
galactic latitude cutoff.  This indicates that the galactic emission
is not a stationary Gaussian random process, even after the
very bright emission from the Galactic plane has been cut out.
Figure \ref{fig:c_l_vs_b.eps} shows the Hauser-Peebles \markcite{HP73}
(1970) estimate of the
power spectrum of the DIRBE 240 $\mu$m map with various cutoff
latitudes.  The lowest latitude cutoff is labeled DC for
\designercut\ -- this is the same part of the sky used for the
analysis of the 4 year DMR maps (Bennett \etal\markcite{BBGHJ96} 1996).
This is basically a $|b| > 20^\circ$
cut with some further exclusions in Taurus-Orion and Scorpius-Ophiucus,
and uses 63\% of the sky which is the area given by a straight
% assumed 3881 pixels - TBD!!
$|b| > 22^\circ$ cut.
The power spectrum in this \designercut\ region is very similar in magnitude 
to the $|b| > 30^\circ$ cut, 
% referee comment `p5'
even though the 63\% sky coverage of the \designercut\ is quite a bit larger
than the 50\% sky coverage in the $|b| > 30^\circ$ cut.
% which indicates that the goal of minimizing
% galactic contamination while maximizing sky coverage was achieved.
However, the $|b| > 40^\circ$ and especially the $|b| > 48.6^\circ$
cuts give power spectra that are lower still but at a big cost in sky 
coverage (36\% and 25\% coverage respectively).

\section{High frequency power spectrum}

The Wright \etal \markcite{WSBL94} (1994) method of finding power spectra
is very slow for large $\ell$'s, so an alternative approach using the 
fast Fourier transform has been used for $\ell \geq 9$.  The equal area
``polar eyeball'' projection used in Figures \ref{fig:band8sd.eps} and
\ref{fig:band10sd.eps} is used to map one galactic hemisphere onto a
square with $512 \times 512$ pixels.  This map is windowed with the
function
\be
W = \cases{\exp\left(\frac{-(1-\sin|b|)}{1-4(1-\sin|b|)}\right) &
	for $|\sin b| > 0.75$;\cr
	0 & otherwise\cr}
\ee
% A \pi r^2 = solid angle = 2\pi (1-\sin b)
% A = 2/256^2
% (r/256)^2 = (1-\sin|b|), so (r/127.5)^2 \approx 4(1-\sin|b|)
which vanishes smoothly at 
$\sin|b| = 0.75$ or $|b| = 48.6^\circ$.  Thus only the high latitude
25\% of sky is used.

\begin{figure}[tb]
\plotone{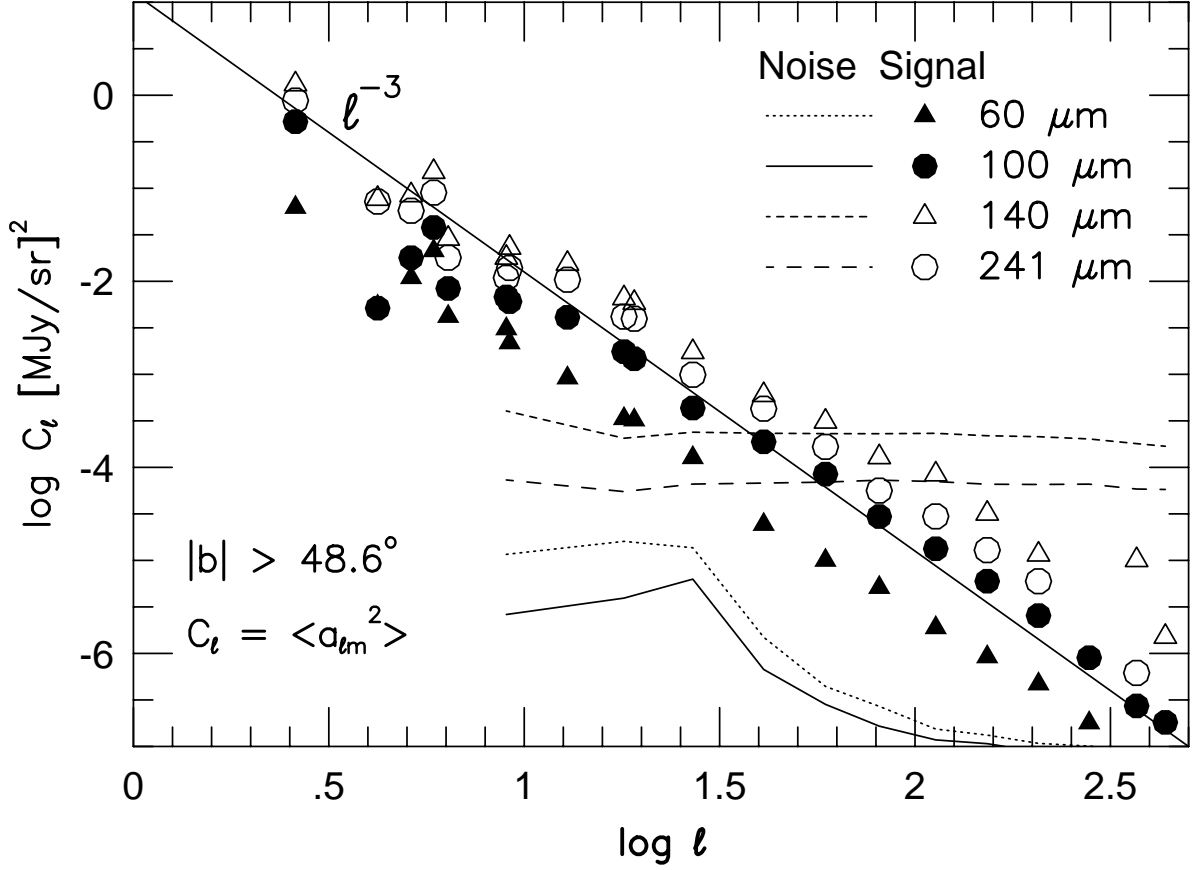}
\caption{Power spectra of the DIRBE 60, 100,
140 and 240 $\mu$m maps in $|b| > 49^\circ$.
% referee comment `Fig 4'
The noise spectra for 140 and 240 $\mu$m are flat, as expected for a detector
noise dominated map.  The noise spectra for 60 and 100 $\mu$m have a pronouced
shoulder at $\ell = 26$, which is due to residual  zodiacal light modeling
error in alternate weeks.\label{fig:hilat.eps}}
\end{figure}

A 2-D fast Fourier transform is then applied to this windowed map.
Since the 512 pixel size of the image covered a hemisphere or
$180^\circ$, the frequency spacing in the output nominally corresponds
to $\Delta \ell = 2$.  However, the use of an equal area projection 
instead of an azimuthal equidistant projection expands the central
portion of the map by a factor $\pi/\sqrt{8}$, so the actual
output step corresponds to $\Delta \ell = 2\sqrt{8}/\pi = 1.8$.
The power spectra of the Northern and Southern polar caps were added
and then summed into rings of constant $|k|$.
Finally the power spectrum of the difference map was subtracted from
the power spectrum of the sum map to give the power spectrum of the sky.
Figure \ref{fig:hilat.eps} shows the resulting power spectra for 
DIRBE bands 7-10: 60, 100, 140 and 240 $\mu$m.  
Points with $\ell \leq 19$ come from
the Wright \etal \markcite{WSBL94} (1994) Hauser-Peebles method applied
to the sky with $|b| > 48.6^\circ$, while the FFT method gives points
with $\ell \geq 9$.  The agreement in the overlap region is good.

\section {Comparison with IRAS}

The power spectra of several cirrus clouds
% referee comment `p6'
observed by IRAS
are given by 
Gautier \etal \markcite{GBPP92} (1992) in the form
\be
P(\rho) = P(0) (\rho/\rho_\circ)^\alpha
\ee
with $\rho_\circ = 0.01$~cycles/arcmin and $P(0)$ ranging from 
$4 \times 10^7$ to $4 \times 10^{10}$ Jy$^2$/sr.  In order to convert
this to $C_\ell$'s I first note that $\rho_\circ$ corresponds to
$\ell = 216$ since $\ell$ is the number of cycles in $2\pi$ radians
$= 21600^\prime$.  The variance of the sky in a band of frequencies
of width $d\rho = \rho_\circ$ centered on $\rho_\circ$ is 
$2\pi\rho_\circ^2 P(0) = 7425 \; \mbox{sr}^{-1} P(0)$.
The same variance computed from $C_\ell$ is
$(4\pi)^{-1}\ell(2\ell+1)C_\ell$ at $\ell = 216$.  These variances are
identical so
\be
C_\ell = 4\pi^2 (\rho/\ell)^2 [1+(2\ell)^{-1}] P(\rho).
\ee
The factor $(\rho/\ell)^2 = (4\pi^2 \mbox{sr})^{-1}$ so the net result is
that $C_\ell$ in (MJy/sr)$^2$ is numerically equal to $P(\rho)$ in
MJy$^2$/sr except for the negligible $[1+(2\ell)^{-1}]$ correction factor.

I estimated the values of $P(0)$ at $\rho_\circ = 0.01$ cycles/arcmin
for the power spectra shown in Figure \ref{fig:hilat.eps} 
using $P(0) = \mbox{median}\{(\ell/216)^3 C_\ell\}$.  The values obtained
are $P(0) = 2.2 \times 10^5,\; 1.0 \times 10^6,\; 3.6 \times 10^6$, and
$2.1 \times 10^6$ Jy$^2$/sr at 60,100, 140, and 240 $\mu$m.
The value at 100 $\mu$m is considerably below the range seen by
Gautier \etal \markcite{GBPP92} (1992) in cirrus clouds, but it is higher
than their extrapolation to dim cirrus.  Since most of the sky with
$|b| > 48.6^\circ$ is quite dim, the IRAS and DIRBE
power spectra agree reasonably well, with the DIRBE measurement for
$|b| > 48.6^\circ$ only $2.5 \times$ the
Gautier \etal \markcite{GBPP92} (1992) estimate for $|b| > 50^\circ$.

\begin{figure}[tb]
\plotone{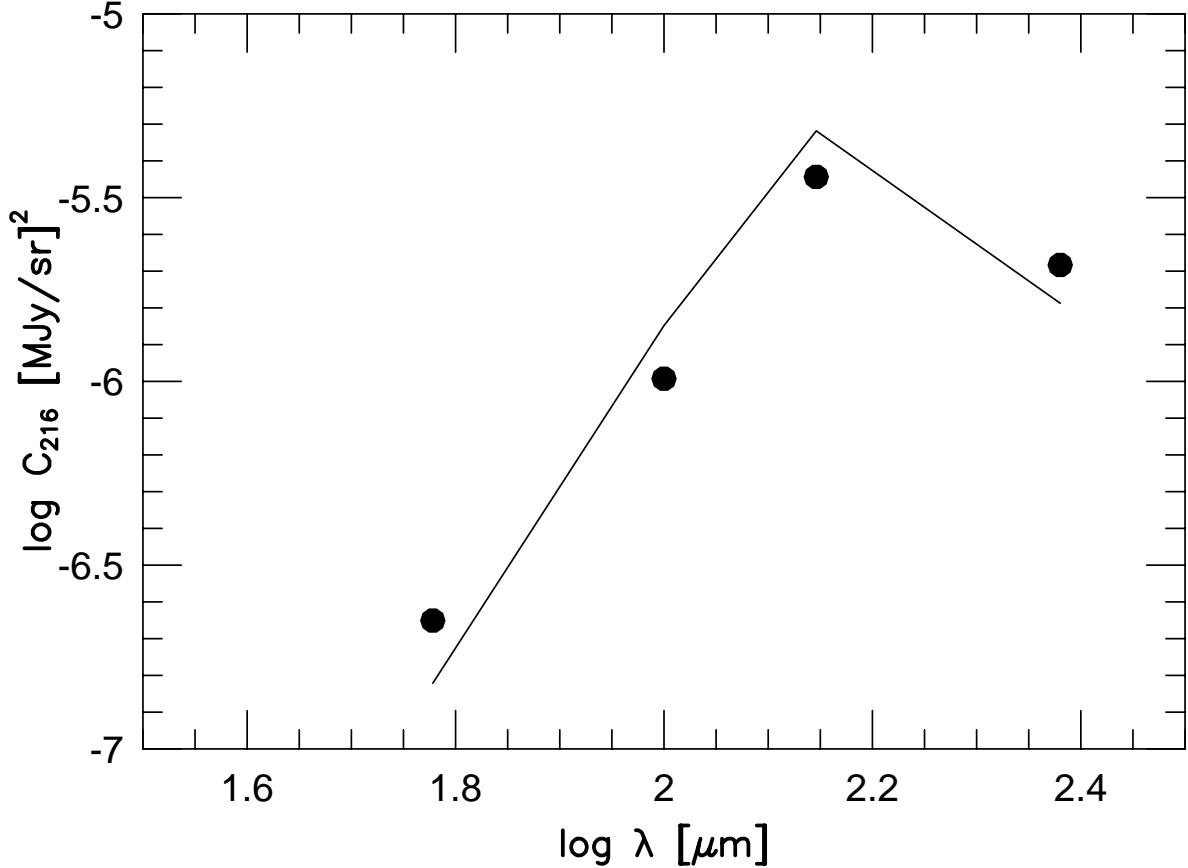}
\figcaption[p0vsnu.eps]{Power spectrum amplitude $C_{216}$ {\it vs.}
wavelength (points) compared to the square of the galactic 
flux.\label{fig:p0vsnu.eps}}
\end{figure}

If the cirrus clouds have the same geometry at all wavelengths, then
the power spectrum should scale with frequency like 
$C_{216} \propto P(0) \propto F_\nu^2$ where $F_\nu$ is the flux of a 
typical cloud.  Using the flux of the Milky Way for $F_\nu$, I find that
\be
C_{216} = P(0) = 10^{-9.74} F_\nu^2.
\ee
This scaling is shown in Figure \ref{fig:p0vsnu.eps}.

%                                          1995          1996
% Enter file name:c_ell.7	P(0) =   2.0859918E-07  2.235E-7
% Enter file name:c_ell.8	P(0) =   1.1590400E-06  1.017E-6
% Enter file name:c_ell.9	P(0) =   3.7887612E-06  3.601E-6
% Enter file name:c_ell.10	P(0) =   2.1633011E-06  2.074E-6

% Table 2 of Gautier etal
% alpha log res ratio  log sep ratio   (E_o double)
%  -2.6    0.00            0.20          -3.159
%  -2.6    0.00            0.40          -2.989
%  -3.2    0.00            0.20          -2.956
%  -3.2    0.00            0.40          -2.745
%  -2.6    0.00            0.30          -3.074
%  -3.2    0.00            0.30          -2.851
%  -3.0    0.00            0.30          -2.925
% E_\circ + [1-(1/2)\alpha]log(d) = 2.1E-4
% sqrt(100^{-3}*P(0)) = sqrt(1.16) = 1.08
% sigma for SIRTF = 0.23 mJy
% ISO is (85/60)^2.5 = 2.4 times worse = 0.54 mJy
% 2.5 m at 240 micron -> 0.396'

The observed power spectrum can be used to estimate the 
typical cirrus contribution to the noise using the tables in
Gautier \etal \markcite{GBPP92} (1992).
With an aperture of $1.2\lambda/D$, and subtracting two reference apertures
each two aperture diameters away, the expected cirrus noise at 100 $\mu$m is
0.23 mJy for {\sl SIRTF} and 0.54 mJy for {\sl ISO}.  
These estimates are 40\% smaller than the estimates in 
Gautier \etal \markcite{GBPP92} (1992).
My calculation requires a large extrapolation in angular scale from the
$0.7^\circ$ DIRBE beam to the $0.5^\prime$ {\sl SIRTF} beam, 
but no extrapolation to the typical high latitude sky.  
% referee comment `p7'
The cirrus noise at 240 $\mu$m in a 3 meter submillimeter telescope
such the planned {\sl FIRST} mission
using the same double subtraction scheme is 0.11 mJy.
The Milky Way would give a flux 8 times greater than this
cirrus noise at $z = 0.7$ for $H_\circ = 50$~km/sec/Mpc and $\Omega = 1$.

\section{Discussion}

The angular power spectra of the high galactic latitude infrared sky
observed by DIRBE show the same power law behavior seen in the
IRAS maps: $C_\ell \propto \ell^{-3}$.  This is one power of $\ell$
steeper than the power spectrum expected for the cosmic microwave background,
so the problem of galactic contamination in moderate resolution
experiments will be smaller than the galactic contamination in the
{\sl COBE}\/ DMR maps.  Thus proposed new CMB anisotropy experiments
that observe with $\approx 0.5^\circ$ beams will not be ruined by
contamination from galactic dust.

\acknowledgements

The DIRBE data products are the result of dedicated work by hundreds of
scientists and engineers who made the {\sl COBE} project a success.
I have had many useful discussions with W. Reach, T. Kelsall, M. Hauser
and other DIRBE team members about zodiacal light modeling.
However, the particular zodiacal light model used in this paper and 
especially its errors shown in Figure \ref{fig:band8sd.eps} 
are entirely my own.

\appendix

\section{Appendix: Zodiacal Light Modeling}
\label{app.zodi}
The models used in this paper are examples of physical models that
integrate a volume emissivity along the the line of sight specified by
unit vector $\hat{n}$:
\be
I_i(\hat{n},t) = \int \rho(\vec{r}(s)) \left[
\kappa_i  D_i(T(R)) + 
\sigma_i \Phi(\mu) D_i(T_{\sun}) R^{-2} \right] ds
\ee
where $R = \vert\vec{r}\vert$,
$\kappa_i$ is a coefficient giving the efficiency in the $i^{th}$
DIRBE band
of the interplanetary
dust grains for thermal emission, while $\sigma_i$ gives their scattering
efficiency.  $\Phi(\mu)$ is the phase function for scattering in terms
of $\mu$, the cosine of the scattering angle.
$D_i(T)$ is the integral of the Planck function through the DIRBE
filter response functions.
The position vector $\vec{r}$ is given by
\be
\vec{r}(s) = \vec{x}_{\earth}(t) + \hat{n}s
\label{eqn:earth}
\ee
where $\vec{x}_{\earth}(t)$ is the position of the Earth in its orbit 
around the Sun.
% referee comment re referring to Tables
The density  $\rho(\vec{r})$ and
temperature $T(R)$ of the interplanetary dust cloud, and the scattering
phase function $\Phi(\mu)$ are all represented by functions with adjustable
parameters $p_i$ having values listed in Table \ref{tab.zodicloud}.
Scaled values of
the grain efficiency factors $\kappa_i$ and $\sigma_i$ are listed in Table
\ref{tab.zodiemiss} with scalings given by
Equations \ref{eq:sigscale} and \ref{eq:kapscale}.
The values of all the parameters are determined by assuming that all of 
the observed time variation
in the DIRBE signal for a given pixel $\hat{n}$ is due to the motion of the
Earth around the Sun described in Equation \ref{eqn:earth}.

However, the symmetry axis of the zodiacal cloud is not the ecliptic
pole but is both tilted and offset
(Dermott \etal\markcite{DNBH84} 1994).  
The tilt is parameterized by the pole unit vector
given by
\be
\hat{z}_C = \frac{(p_6,p_7,10)}{\sqrt{100+p_6^2+p_7^2}}
\ee
The inclination $i$ with respect to the cloud is specified by
\be
\sin i = \frac{\hat{z}_C\cdot\vec{r}}{R}.
\ee
The density contours in the cloud are also not centered on the Sun.
An offset radius is defined as
\be
R_c = R + \vec{o}_C\cdot\vec{r}
\ee
where
\be
\vec{o}_C = (p_8/10, p_9/10, 0).
\ee

\begin{table}[tbp]
\begin{center}
\begin{tabular}{ccl}
\hline
Parameter &
Value & 
Description\\ 
\hline
$p_1$ &        1.2186 & radial density exponent \nl
$p_2$ &        0.4451 & radial temperature exponent \nl
$p_3$ &        3.6122 & vertical ``scale height'' \nl
$p_4$ &        0.9285 & vertical density exponent \nl
$p_5$ &       -1.4766 & $\ln(\sin i)$ at break \nl
$p_6$ &        0.3705 & $10\times$ cloud pole $x$ component \nl
$p_7$ &       -0.0736 & $10\times$ cloud pole $y$ component \nl
$p_8$ &       -0.0235 & $10\times$ cloud offset $x$ component \nl
$p_9$ &       -0.0081 & $10\times$ cloud offset $y$ component  \nl
$p_{10}$ &     0.7548 & $10\times$ density contrast of Dermott ring\nl
$p_{11}$ &     5.5301 & $\ln(T_B)$, band temperature at $R=1$ \nl
$p_{12}$ &     0.7727 & scale factor for bands \nl
$p_{13}$ &     0.4284 & the ``dimple'' in Dermott's ring \nl
$p_{14}$ &    27.7741 & vertical scale for Dermott's ring \nl
$p_{15}$ &    -0.0251 & spherical term in vertical density \nl
$p_{16}$ &     0.0249 & $\sin^2 i$ term in vertical density \nl
$p_{17}$ &    -0.0456 & additional density at $|\sin i| \approx 0.5$ \nl
$p_{18}$ &    -0.1276 & additional density at $|\sin i| \approx 0.25$ \nl
$p_{19}$ &    -0.0103 & additional density at $|\sin i| \approx 0.17$ \nl
$p_{20}$ &    -0.3133 & phase function linear coefficient \nl
$p_{21}$ &     0.5749 & phase function quadratic coefficient \nl
\hline
\end{tabular}
\caption{Diffuse Cloud Parameters \label{tab.zodicloud}}
\end{center}
\end{table}

The simplest models for the zodiacal cloud assume
\be
\rho(\vec{r}) = \rho_\circ f(i) R_c^{-p_1}
\ee
where $f(i)$ specifies the heliocentric inclination dependence of a
fan shaped cloud and $p_1$ is the exponent of a power law density dependence.
The density scaling $\rho_\circ$ is fixed at unity and the overall strength 
of the zodiacal signal is determined by the $\kappa_i$'s and the $\sigma_i$'s.
The temperature dependence is also given by a power law:
\be
T(r) = T_\circ R^{-p_2}
\ee
The temperature was fixed at $R = 1$~au at 280~K because of a strong degeneracy
between the $\kappa_i$'s and $T_\circ$.
The inclination dependence $f(i)$ is given by
\bea
f(i) & = & \exp(-p_3\times Z^{p_4}) + p_{15} 
   +  p_{16}\times \sin^2i + 
p_{17}\times 4 \sin^2i\,\exp(-4\sin^2i) \nonumber\\
& + &
p_{18}\times 16 \sin^2i\,\exp(-16\sin^2i) +
p_{19}\times 36 \sin^2i\,\exp(-36\sin^2i))
\eea
where $Z$ is ``rounded'' absolute value of $\sin i$:
\be
Z = \cases{\vert\sin i\vert - \case{1}{2} S & for $\vert\sin i\vert > S$;\cr
           \case{1}{2} \sin^2 i /S & for $\vert\sin i\vert \leq S$.\cr}
\ee
The break point is given by $S = \exp(p_5)$.

The ring of particles resonantly trapped by gravitational perturbations from
the Earth 
(Dermott \etal\markcite{DJXGL94}\ 1994; 
Reach \etal\markcite{RFWHK95} 1995) 
is modeled
using coordinates in a frame rotating to follow $L_{\earth}$,
the mean orbital longitude of the Earth:
\bea
x_D & = & x \cos L_{\earth} + y \sin L_{\earth}\nonumber\\
y_D & = & y \cos L_{\earth} - x \sin L_{\earth}\nonumber\\
z_D & = & z
\eea
Then the following terms are defined:
\bea
L_D & = & \left|\mbox{atan2}(y_D,x_D) + 0.25\right|\\
A & = & \cases{\cos(8\pi L_D/3) & for $L_D < \case{3}{8}$;\cr
               (\cos(8\pi L_D/3)-1)/2 & 
		for $\case{3}{8} <L_D < \case{3}{4}$;\cr
		0 & otherwise.\cr}\\
D & = & \exp\left[-56.5\left(\sqrt{x_D^2+y_D^2}-1.133+
0.133\,p_{13}\,\exp(-4 L_D^2)
\right)^2 - \frac{p_{14}z_D^2}{R^2} \right]
\eea

The final form for the density is
\be
\rho(\vec{r}) = \frac{R}{R_c} f(i) R_c^{-p_1}
\left(1 + \frac{p_{10}D(1+A)}{10}\right).
\ee

\begin{deluxetable}{ccl}
\tablecaption{Scattering and Emission Efficiencies \label{tab.zodiemiss}}
\tablehead{
\colhead{Parameter} &
\colhead{Value} & 
\colhead{Description} 
}
\startdata
$p_{22}$ &     0.6864 & $\sigma_1$ scaling \nl
$p_{23}$ &     0.8362 & $\sigma_2$ scaling \nl
$p_{24}$ &     0.6346 & $\sigma_3$ scaling \nl
$p_{25}$ &     2.2829 & $\kappa_{ 3}$ scaling \nl
$p_{26}$ &     1.4175 & $\kappa_{ 4}$ scaling \nl
$p_{27}$ &     1.2422 & $\kappa_{ 5}$ scaling \nl
$p_{28}$ &     1.2996 & $\kappa_{ 6}$ scaling \nl
$p_{29}$ &     0.8788 & $\kappa_{ 7}$ scaling \nl
$p_{30}$ &     0.7646 & $\kappa_{ 8}$ scaling \nl
$p_{31}$ &     0.8243 & $\kappa_{ 9}$ scaling \nl
$p_{32}$ &     0.6254 & $\kappa_{10}$ scaling \nl
\enddata
\end{deluxetable}

\begin{deluxetable}{ccl}
\tablecaption{IRAS Band Parameters \label{tab.bandpar}}
\tablehead{
\colhead{Parameter} &
\colhead{Value} & 
\colhead{Description} 
}
\startdata
$q_1$ &   1.3849  & $10\times(\sin i)_{max}$ for Band 1 \nl
$q_2$ &   0.2743  & Band 1 normalization \nl
$q_3$ &   0.2807  & $10\times(\sin i)_{max}$ for Band 2 \nl
$q_4$ &   0.4407  & Band 2 normalization \nl
$q_5$ &   0.1735  &  $10\times$ band pole $x$ component \nl
$q_6$ &  -0.2088  & $10\times$ band pole $y$ component \nl
$q_7$ &  -1.5723  & $10\times$ band offset $x$ component \nl
$q_8$ &  -0.2225  & $10\times$ band offset $y$ component \nl
$R_1$ & 3.14 & Outer radius for Band 1 \nl
$R_2$ & 3.02 & Outer radius for Band 2 \nl
\enddata
\end{deluxetable}

The phase function for scattering is given by
\be
\Phi(\mu) = \exp(-p_{20}\mu + p_{21} \mu^2).
\ee
The values of $\kappa_i$ and $\sigma_i$ are also adjustable parameters.
Only bands 1--3 ($J$, $K$ and $L$) have non-zero $\sigma_i$'s, with
\be
\sigma_i = 1.8 \, p_{21+i} \times 10^{-13}
\label{eq:sigscale}
\ee
and only bands 3--10 have non-zero $\kappa_i$'s with
\be
\kappa_i = 10^{-7} p_{22+i}.
\label{eq:kapscale}
\ee

However, this model does not include the IRAS bands 
(Low \etal\markcite{LYBGB84} 1984)
The bands are described by more parametrized functions with parameter values
given in Table \ref{tab.bandpar}.
So the following term is added to the intensity:
\be
I_B = p_{12} \int \rho_B(\vec{r}(s)) \left[
\kappa_i  D_i(T_B(R)) + 
\sigma_i \Phi(\mu) D_i(T_{\sun}) R^{-2} \right] ds
\ee
with
\be
T_B(R) = \frac{\exp(p_{11})}{\sqrt{R}}
\ee

The offset and tilt of the bands are given by
\be
\hat{z}_B = \frac{(q_5,q_6,10)}{\sqrt{100+q_5^2+q_6^2}}
\ee
where $q_n$ will denote adjustable parameters in the models.
The inclination $i_B$ with respect to the bands is specified by
\be
\sin i_B = \frac{\hat{z}_B\cdot\vec{r}}{R}.
\ee
The density contours in the bands are also not centered on the Sun.
The offset radius is defined as
\be
R_B = R + \vec{o}_B\cdot\vec{r}
\ee
where
\be
\vec{o}_B = (q_7/10, q_8/10, 0).
\ee
The density in the bands is given by
\be
\rho_B = \frac{R}{R_B^2} \sum_{j=1}^2 {q_{2j} \left[
         \cases{
         \cosh\left(\frac{1.72 \vert \sin i_B \vert}{\mbox{$q_{2j-1}$}}\right)
         & for $\vert \sin i_B \vert < q_{2j-1}$ and $R_B < R_j$;\cr
       0 & otherwise.}\right]}
\ee
The functional form $\cosh(1.72 x)$ was chosen instead of
$(1-x^2)^{-1/2}$ to avoid numerical problems when integrating
through the cusps.  Note that the coefficient is chosen so that
$\int_{-1}^{+1} \cosh(1.72 x) dx = \pi = \int_{-1}^{+1} (1-x^2)^{-1/2} dx$.
The outer radii are taken from
Jones \& Rowan-Robinson \markcite{JR93} (1993).

The band parameters $\{q_j\}$ and the diffuse cloud parameters $\{p_i\}$
are optimized in alternating runs.  When optimizing the band parameters
a set of normal points emphasizing the ecliptic is used.
When optimizing the diffuse cloud parameters a separate set of normal
points emphasizing the ecliptic poles is used.

A second model that used a slightly different treatment of the Dermott
ring and clump was also tried.  An additional parameter $P_1$ describes 
the radial extent
of the ring:
\be
D = \exp\left[-56.5 P_1 
\left(\sqrt{x_D^2+y_D^2}-1.133+0.133 p_{13}\exp(-4 L_D^2)
\right)^2 - \frac{p_{14}z_D^2}{R^2} \right]
\ee
A second additional parameter $P_2$ gives a separate adjustment
for the density of the ``clump'' trailing the Earth.  The
density of the cloud is given by
\be
\rho(\vec{r}) = \frac{R}{R_c} f(i) R_c^{-p_1}
\left(1 + \frac{p_{10}D+P_2 D A)}{10}\right)
\ee
Fitting for these parameters gives $P_2 > p_{10}$ which implies a
negative density in the ring+clump system, which seemed unphysical,
so the simpler model described above was developed.

\end{document}